\documentclass[aps,preprint,superscriptaddress,amsmath,amssymb]{revtex4}
\input psfig.sty
\newcommand{\Op}[1]{{\boldsymbol{\mathrm{\hat{#1}}}}}
\newcommand{\Fkt}[1]{{\mathrm {#1}}}
\newcommand{\sub}[1]{_{\mathrm {#1}}}

\usepackage{graphicx}% Include figure files
\usepackage{dcolumn}% Align table columns on decimal point
\usepackage{bm}% bold math
\begin{document}
%\preprint{}

%\begin{frontmatter}


\title{Decoherence control: A feedback mechanism based on hamiltonian tracking} 

%%%%%%%%%%%%%%%%%%%%%%%%%%%%%%%%%%%%%%%%%%%%%%%%

\author{Gil Katz}
%\email{gkatz@chem.northwestern.edu}
\affiliation{Department of Chemistry, Northwestern University, Evanston, Illinois 60208-3113}
\author{Mark A. Ratner}
%\email{ratner@chem.northwestern.edu}
\affiliation{Department of Chemistry, Northwestern University, Evanston, Illinois 60208-3113}
\author{Ronnie Kosloff}
%\email{ronnie@fh.huji.ac.il}
\affiliation{Fritz Haber Research Center for Molecular Dynamics,\\
Hebrew University of Jerusalem, Jerusalem, 91904, Israel}

%%%%%%%%%%%%%%%%%%%%%%%%%%%%%%%%%%%%%%%%%%%%%%%%%%%%%%%%%%%%%5555

%%%%%%%%%%%%%%%%%%%%%%%%%%%%%%%%%%%%%%%%%%%%%%%%%%%%%%%%%%%%%%%%%%%%%%%

\begin{abstract}
Enviroment - caused dissipation disrupts the hamiltonian evolution of all quantum systems 
not fully isolated from any bath. We propose and examine a feedback-control scheme to eliminate
such dissipation, by tracking the free hamiltonian evolution.
We determine a driving-field that maximizes the projection of the actual molecular 
system onto the freely propagated one.   
The evolution of a model two level system in a dephasing bath is followed, 
and the driving field that overcomes the decoherence is calculated.
An implementation of the scheme in the laboratory using feedback control is suggested.

\\

 \end{abstract}

\date{\today}
%\received{}
%\accepted{}

%\begin{keyword}
% markovian decoherence process \sep
% tracking control\sep
% PACS 82.50.Nd \sep 02.30.Yy \sep
%\end{keyword}
%\end{frontmatter}
\maketitle

% main text

%\section{Introduction}
%\label{sec:intro}

The Schr{\"{o}}dinger equation describes time evolution of isolated quantum systems. Any real system, unless very carefully isolated, will 
decohere due to bath interactions\cite{Fraval,Mukamel1,Mukamel2,tannorDC1,tannorDC2,Rossky1,Rossky2,Zurek1,Zurek2,Zurek3}. 

The approach to decoherence correction taken here derives from tracking control theory.
Tracking control originated from the idea of changing the local field in order to follow a 
reference state \cite{rabitz4,rabitz6,schwentner1,schwentner2}.
Tracking can be considered as local control with a time dependent target.Since generation of interference pathways is  
a global process in time, an intervention at one instant influences the subsequent 
evolution and can interfere at later times with another intervention.

The basic idea here is to use a freely propagating molecule as a tracking target
for an identical molecule in a dissipative environment. The perfect tracking field will then null the 
dissipative forces and protect the molecule from decoherence. The question to be addressed is how
can an external field, that generates a unitary transformation, protect the system
against non unitary dissipative forces?  A computational model of
an on-line, dynamic, decoherence control method is explored.
The method is established by simultaneously following the evolution in time of 
a system with and without the bath, thus using the 
overlap between the two systems to construct a correcting electric field which 
is directly applied back on the system.      
The method is explored for the fastest dissipative mechanism, environment-induced electronic quenching 
from an excited state to 
a lower state (the approximate decay time in solid 
state systems is less than 50 fsec). The coherence  
time of the system is extended by an order of magnitude using the simplest 
tracking method.

Measuring the overlap between the controlled and the target system also underlies 
a proposed experimental approach. This observable can then be employed as a basis for
a learning loop used to optimize a control field.

%\section{Theory}
%\label{sec:theory}

The molecular system is described by the density operator $\Op \rho$, a function of nuclear and
electronic degrees of freedom. The reference or target 
system $\Op \rho_{tar}$, evolves freely under its Hamiltonian ${\Op H}_0$:

\begin{equation}
{ { \Op \dot  \rho}_{tar}}  = -i [{\Op H_0},{\Op \rho_{tar}}] 
%{ {  \Op \dot  \rho}}  = -i [{\Op H_0},{\Op \rho}] 
\label{eq:H0}
\end{equation}

where ${\hbar \equiv 1}$ is chosen.
The subjected system $\Op \rho_{S}$ is coupled to a bath and therefore evolves according to the dynamics
of an open quantum system \cite{lindblad}: 
\begin{equation}
{\dot{\Op  \rho}_{S}}  = -i[\Op H_0,\Op \rho_{S}]+ 
{\cal L}_D({\Op \rho}_{S})~~,
\label{eq:Hsb}
\end{equation}

Here ${\cal L}_D$ describes the bath-induced decoherence dynamics, conveniently cast into  the Lindblad semigroup form:
\begin{equation}
{\cal L}\sub{D}(\Op \rho\sub{S})= \sum_j \left( \Op F_j \Op
\rho\sub{S} \Op F_j^{\dagger}- \frac{1}{2}\{\Op F_j \Op F_j^{\dagger},\Op
\rho\sub{S} \} \right) ~~,
\label{eq:lindbladOP}
\end{equation}
where $\{\Op A, \Op B\}=\Op A \Op B+\Op B \Op A$ is the anti commutator and $\Op F $ 
are operators from the Hilbert space of the system. 
The nature of the bath interaction determines the form of the Lindblad operators $\Op F$.
The choice of $\Op F $ determines the dissipative model considered.

Our purpose is to control the system coupled to the bath 
and force it to follow as closely as possible the dynamics of the freely evolving 
reference system, $\Op \rho_{tar}$. The external control field  $\epsilon (t)$ is coupled to the system through the transition dipole of the molecule. 
The control Hamiltonian becomes, ${\Op H_c} =\Op \mu \epsilon(t)$ leading to the dynamics of the controlled system $\Op \rho_C$:
\begin{equation}
 \dot {\Op \rho}_{C} = -i[{\Op H}_0,\Op \rho_C]
-i[{\Op H}_c,{\Op \rho}_C]+ {\cal L}_D({\Op \rho}_{C})~~.
\label{eq:Hc}
\end{equation} 
The challenge of decoherence control is to find the field, $\epsilon(t)$, 
that induces the dynamics of the system $\Op \rho_C(t)$ to be as close as possible to $\Op \rho_{tar}$. 
This is attacked by maximizing the overlap functional $ J(t)$ between the target
and controlled states:
\begin{equation}
J(t)= \left ( \Op \rho_C \cdot \Op \rho_{tar} \right ) \equiv  
Tr\{ \Op \rho_C ~\Op  \rho_{tar}\} 
= \langle \Op \rho_{tar} \rangle 
\end{equation}
$J(t)$ can be interpreted as the expectation of the target density operator in the controlled state.
To achieve this target the control field should increase the expectation of the target operator.
The Heisenberg equation of motion for the observable $\langle \Op \rho_{tar} \rangle $ generated
by the control Hamiltonian becomes: 
\begin{eqnarray}
 \frac{d}{dt} \langle { \Op \rho}_{tar} \rangle  ~=~  
- i {\langle [\Op H_c,{ \Op {\rho_{tar}}]} \rangle} ~=~ \\
- i \epsilon(t)  Tr\{ \Op \mu  \Op {\rho_{tar}} \Op {\rho_{C}} -
\Op {\rho_{tar}} \Op \mu  \Op {\rho_{C}} \}~~.
\end{eqnarray}
The control field is constructed by requiring maximal $ {J(t)}$, so that 
$ \frac{d}{dt} \langle \bf{\Op \rho}_{tar} \rangle \geq 0 $ at any instant,
leading to:
\begin{equation}
\epsilon (t)= -i K~ Tr\{ \Op {\rho_{C}} \Op \mu  \Op {\rho_{tar}}  -
\Op {\rho_{tar}} \Op \mu  \Op {\rho_{C}} \}^*~~,
\label{eq:contfield}
\end{equation}
where $K=K(t)$ is a positive envelope function with dimension $\frac {energy}{(dipole)^2}$ . From Eq. (\ref{eq:contfield}) it is clear
that the correcting field $\epsilon (t)=0$ if $\Op {\rho_{C}}$ approaches $\Op {\rho_{tar}}$.

%\section{Model}
%\label{sec:model}

We adopt a molecular model system with two electronic states
described by the density operator:
\begin{equation}
 \Op{\rho} ~~=~~  \begin{pmatrix}
\Op{\rho}\sub{e}                  &   \Op{ \rho}\sub{eg} \\ 
\Op \rho \sub{ge} &       \Op{\rho}\sub{g}
\end{pmatrix}~,
\label{eq:rho}
\end{equation}
where the indices $g$ and $e$ designate the ground and excited electronic states and 
the sub matrices are functions of the nuclear coordinates.
The Hamiltonian of the system has the form:
\begin{equation}
\Op{H_0} =  \begin{pmatrix}
\Op{H}\sub{e}                  &   \Op {V}_{eg} \\ 
\Op {V}_{ge} &       \Op{H}\sub{g}
\end{pmatrix}~,
\label{eq:Hsystem}
\end{equation}
with $\Op{H}\sub{g/e} = \Op{T} +\Op V\sub{g/e}$.
$\Op{T}$ is the kinetic energy operator, $\Op V\sub{g}$ and $\Op V\sub{e}$ 
are the potential energy operators on the ground and excited electronic 
states and $\Op {V}_{eg}$ is the nonadiabatic coupling potential.
The control Hamiltonian is chosen as:
\begin{equation}
\Op{H}\sub{c} =  \begin{pmatrix}
 0 & - \Op {\mu^{\dagger}}_i\epsilon(t)  \\ 
- \Op \mu \epsilon(t) &  0 
\end{pmatrix} ~~,
\end{equation}
where $\Op \mu $ is the coordinate dependent electronic transition dipole element.

The dissipation operator ${\Op F_{q}}$  
is chosen to model fast electronic quenching as 
${\Op F_{q}}= \sqrt{\gamma_{q}}(|e\rangle  \langle g|)$. 
%\begin{equation}
%{\Op{F}\sub{q}}~ =~ {\gamma}^{\frac{1}{2}} \otimes \begin{pmatrix}
% 0 & 0 \\ 
%1 & 0 
%\end{pmatrix} \; ,
%\label{eq:quenching}
%\end{equation}
This choice of dissipative term will lead to a process with a characteristic quenching rate $\gamma$ :
\begin{equation}
 {\cal L}_D (\Op \rho)~~
	  = ~~
   \gamma
    \left (
    \begin{array}{ l r}
 -\Op \rho_{e} & -{\Op \rho_{ge}}^{\dagger}/2\\
 - \Op \rho_{eg}/2 &  \Op \rho_e \\
 \end{array}
  \right ) ~~.
  \end{equation}
  
The control field form Eq. (\ref{eq:contfield}) becomes:
\begin{eqnarray}
\begin{array}{l}
{ \epsilon}(t)  ~~=~~ K Im \left( \right.
Tr_Q \{ {\Op \rho^{ge}_{c}} \Op \mu \Op \rho^{g}_{tar} \} +Tr_Q \{ {\Op \rho^e_c \Op \mu \Op \rho^{ge}_{tar}} \} \\
- Tr_Q \{ \Op \rho^{ge}_{tar} \Op \mu  {\Op \rho^{g}_c} \}  - Tr_Q \{ \Op \rho^{e}_{tar} \Op \mu  \Op {\rho^{ge}_c} \} \\ 
- Tr_Q \{ \Op \rho^{g}_{tar} \Op \mu \Op {\rho^{eg}_c}\} + Tr_Q \{ \Op {\rho^g_c} \Op \mu \Op \rho^{eg}_{tar} \}    \\                                    
- Tr_Q \{ \Op {\rho^{eg}_{tar}}\Op \mu \Op {\rho^{ge}_c} \}  + Tr_Q \{ \Op {\rho^{eg}_c}\Op \mu \Op  \rho^{e}_{tar} \}
\left. \right)  
\end{array}~~,
\label{eq:cfield}
\end{eqnarray}
where $Tr_Q$ is a partial trace over the coordinates.

The computational model is constructed from two electronic states of a diatomic molecule represented by   
two diabatic potential energy surfaces that are approximated by quadratic functions with linear
interstate couplings \cite{koppel84}. Using dimensionless normal coordinates:
\begin{eqnarray}
\begin{array}{lll}
V_{g}(\Op{Q}) &=& -\Delta + \frac{\omega_g^2}{2} {({\Op Q}-Q_g)}^2 \; ,\\ 
V_{e}(\Op{Q}_0) &=& \Delta + \frac{\omega_e^2}{2}  {({\Op Q}-Q_e)}^2 \; ,
\end{array}
\label{eq:pots}
\end{eqnarray}
where $\omega_g$ and $\omega_e$ are the vibrational frequencies of the ground and excited electronic states, 
$2\Delta$ is the adiabatic excitation energy.
$Q_e$ and $Q_g$ are the equilibrium bond lengths of the two electronic states.
%%%%%%%%%%%%%%%%%%%%%%%%%%%%%%%%%%%%%%%%%%%%%%55
\begin{figure}
\includegraphics[scale=0.25]{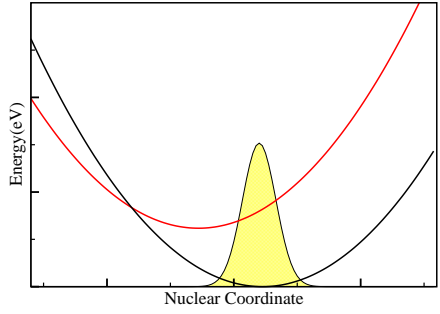}
\caption{\footnotesize The potential energy surface of the molecular model 
and the ground initial state. The parameters describing the model are:
$\omega_g= \omega_e=0.07 ~eV$, $\Delta = .7 ~eV$ ,$V\sub{ge/eg}=0.05 ~eV$, $Q_g=0~~,~~Q_{e}=-0.1 ~ $ 
and $\Op \mu=1.0$.}
%\vspace{0.2cm}
\label{fig:system} 
\end{figure}
%%%%%%%%%%%%%%%%%%%%%%%%%%%%%%%%%%%%%%%%%%%%%%

The time evolution was obtained by solving the time dependent Liouville-von Neumann 
equation. A grid was used for the spatial coordinates and time propagation used the Chebychev scheme
\cite{k144}. The initial ground vibronic state of the system was calculated via a relaxation scheme \cite{k42}.
A pump pulse transfers a significant fraction of the population from the ground to the excited electronic state.
A Gaussian form is chosen for this pulse:
\begin{equation}
\epsilon\sub{pump}(t) = \epsilon_0 \Fkt{e}^{-(t-t_{max})^2/2\sigma_L^2} 
\Fkt{e}^{i \omega_L t} \; , 
\label{eq:pulse}
\end{equation}
The carrier frequency  $\omega_L$ is chosen to match the difference between 
the ground and excited electronic potentials at the minimum of the ground state $Q=0$.
The width of the pulse $\sigma_L$ was adjusted to a FWHM duration  of 12 fsec 
and the amplitude $ \epsilon_0 \cdot \mu =0.228 eV $.

At this point three parallel propagations of the density operator were performed.
The first for $\Op \rho_{tar}$ was carried out only with the Hamiltonian term Eq. (\ref{eq:H0}).
The system  density operator  $\Op \rho_{S}$, was propagated with the dissipative ${\cal L}_D$ included, Eq. (\ref{eq:Hsb}).
The controlled  density operator,  $\Op \rho_{C}$ was propagated with both the dissipative ${\cal L}_D$ and
the correcting field applied terms, Eq. (\ref{eq:Hc}). The correction field was calculated according to
Eq. (\ref{eq:cfield}) at each time step and fed back in the next time step.

%\section{Results and discussion}

\begin{figure}
\includegraphics[scale=0.375]{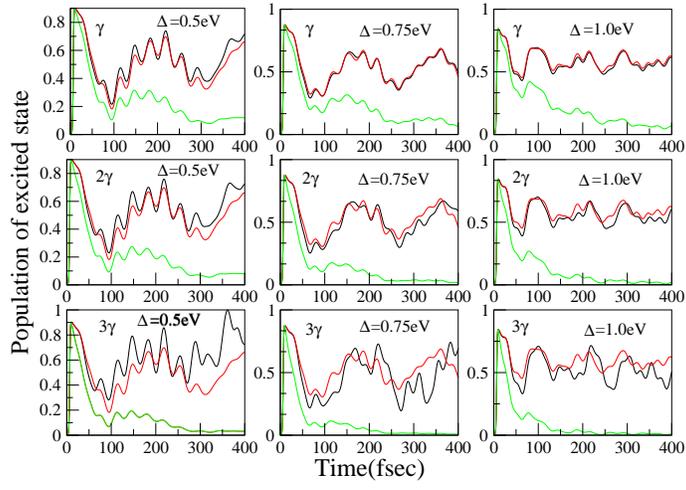}
\caption{\footnotesize The population of the excited state as a function of time 
for the target state $\Op \rho_{tar}$ (red), the controlled density operator $\Op \rho_{C}$(black) 
and the uncontrolled system density operator $\Op \rho_S$ (green)  
for different magnitudes of quenching(columns) and $\Delta$'s(rows).
$\gamma= 0.003 fsec^{-1}$.}
\vspace{0.4cm}
\label{fig:mat}
\end{figure}

\begin{figure}
\includegraphics[scale=0.3]{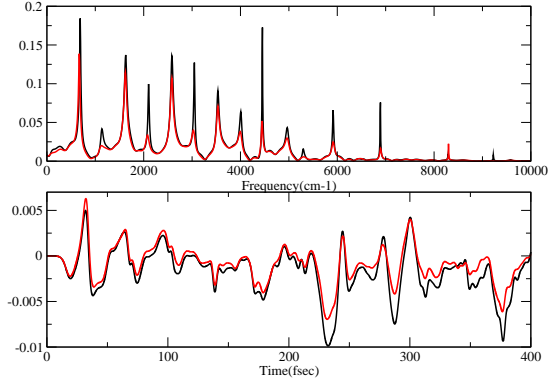}
\caption{\footnotesize  Typical control field in frequency (top left)
and time domain (bottom) for $\gamma=0.003$(red) and $\gamma=0.006 ~$ $fsec^{-1}$(blue). 
%Right - The Wigner distribution of the control field in frequency and time domain.
} 
\label{fig:field}
\vspace{0.3cm}
\end{figure}

The correction scheme was tested for the the system shown in  Fig. \ref{fig:system}.  
The free dynamics represents a complex population oscillation between the two electronic states.
The excited state population is shown in Fig. \ref{fig:mat} 
for the controlled $\Op \rho_c$, uncontrolled $\Op \rho_S$
and target $\Op \rho_{tar}$ for different quenching parameters $\gamma$ and energy gap $\Delta$.
In all cases a fast decay of the excited state is seen for the uncontrolled state.
The controlled state is able to track the target state and maintain the population.
When the quenching increases the controlled system is still 
able to follow the overall dynamics of the target
but not the finer details.  As can be seen in Fig. \ref{fig:mat}, 
the control ability increases with the energy gap $\Delta$. 
The control field $\epsilon$ is shown in Fig. \ref{fig:field} 
in time and frequency. In general the field is
composed of a central frequency corrosponding to the energy gap modulated by the vibrational 
frequency. The actual vibronic lines are missing from the spectrum \cite{k163}.
Wigner plots ( not shown ) show phase locking between the frequency components.  

\begin{figure}
%\vspace{0.2cm}
\includegraphics[scale=0.3]{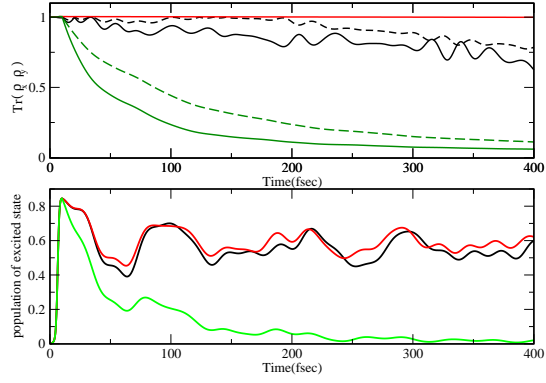}
\caption{\footnotesize 
Top - The purity $Tr \{ \Op \rho^2 \}$ (solid) and the scalar product with the target state 
$\left( \Op \rho \cdot \Op \rho_{tar} \right)$
(dashed) for the target state $\Op \rho_C$ (red) the controlled density 
operator $\Op \rho_{C}$ (black) and the reference uncontrolled density operator $\Op \rho_S$ (green).
Bottom - The population of the excited state, same color codes.}
\vspace{0.4cm}
\label{fig:Scal}
\end{figure}

\begin{figure}
\includegraphics[scale=0.3]{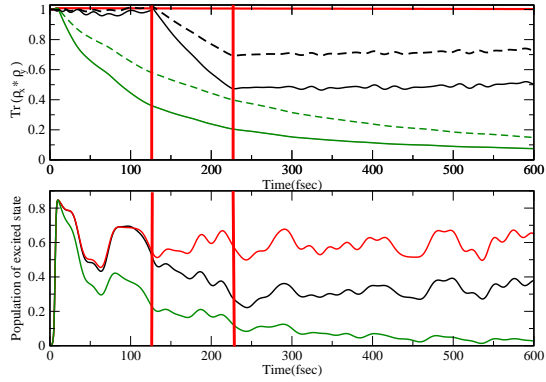}
\caption{\footnotesize 
Turning off the control field.
Top -The purity $Tr \{ \Op \rho^2 \}$ (solid) and the scalar product with the target state 
$\left( \Op \rho \cdot \Op \rho_{tar} \right)$
(dashed) for the target state $\Op \rho_{tar}$ (red) the controlled density 
operator $\Op \rho_{C}$ (black) and the reference uncontrolled density operator $\Op \rho_S$ (green).
The vertical lines indicate  the time interval where the control field is turned off.
Bottom -  The population of the excited state as a function of time, same color codes.}
\vspace{0.3cm}
\label{fig:onoff}
\end{figure}

The changes in purity $Tr \{ \Op \rho^2 \} $ and in the scalar product  with the target state 
$\left( \Op \rho \cdot \Op \rho_{tar} \right)$, show a different viewpoint on the decoherence control Fig. \ref{fig:Scal}.
The uncontrolled state $\Op \rho_S$ undergoes fast exponential decay of both measures in time.
The controlled purity $Tr \{ \Op \rho_C^2 \} $ maintains a high value as does as the scalar product 
$\left( \Op \rho_C \cdot \Op \rho_{tar} \right)$  The general trend is  a linear decay in time. 
The oscillations around this decay and the temporal increase in purity suggest that a cooling mechanism 
is taking place \cite{k163}.
To gain insight on this possibility the correcting field was stopped for 100 fsec and 
resumed again, Fig. \ref{fig:onoff}. As expected, once the control field is turned off
the purity and the scalar product
of the controlled state $\Op \rho_C$ decease sharply. When the controlling field is resumed 
an almost linear increase in both parameters can be observed. 
Such an effect can only be a result of a cooling mechanism
which represents an interplay between the nonunitary dissipation 
and the unitary control field \cite{k163}. It is clear \cite{pritchard} that simple unitary transformation cannot 
induce cooling.

The present results show that the decoherence can be suppressed, extending good overlap 
with the target by an order of magnitude in time. The model calculation establishes 
the principle that a correcting external field can be found which protects the 
system from decoherence. Here we discuss control to overcome fast electronic dephasing; while the rapid modulations
in Fig.  \ref{fig:field} are not currently attainable, the essential behaviour is correct.

The proof of principle then shifts the attention to finding such a field in the laboratory. 
The idea is to use an actual 
freely evolving hamiltonian subsystem as the target
for an identical hamiltonian subsystem subject to a dissipative environment such as in solution 
or on a surface. The correcting feedback should be set such that the 
difference in transient absorption of the two molecules is minimized.

To implement the present scheme in the laboratory the key element is to apply a scalar product between the states
of the reference and controlled systems: $ \left ( \Op \rho_C \cdot \Op \rho_{tar} \right )$ 
(displayed in Fig.\ref{fig:Scal}).  
To make this task possible the two isolated
states should become part of the same quantum system. 
This can be carried out by placing the controlled system $\Op \rho_C$ and the reference 
system $\Op \rho_{tar}$ in two branches of an interferometer.
A pair of twin ultrashort photons which can be created by 
down conversion are split spatially. Then they are directed 
to the two branches of the interferometer interrogating the two systems.
The transmitted photons are then redirected to interfere with each other. 
Any difference between the target and the controlled system will degrade this interference.
The interference signal can then be used as a feedback to generate 
the correcting field \cite{habib}.

\begin{acknowledgments}
We are grateful to the BSF and the NSF for support of this work.
\end{acknowledgments}

\bibliography{Control}

\begin{thebibliography}{21}
\expandafter\ifx\csname natexlab\endcsname\relax\def\natexlab#1{#1}\fi
\expandafter\ifx\csname bibnamefont\endcsname\relax
  \def\bibnamefont#1{#1}\fi
\expandafter\ifx\csname bibfnamefont\endcsname\relax
  \def\bibfnamefont#1{#1}\fi
\expandafter\ifx\csname citenamefont\endcsname\relax
  \def\citenamefont#1{#1}\fi
\expandafter\ifx\csname url\endcsname\relax
  \def\url#1{\texttt{#1}}\fi
\expandafter\ifx\csname urlprefix\endcsname\relax\def\urlprefix{URL }\fi
\providecommand{\bibinfo}[2]{#2}
\providecommand{\eprint}[2][]{\url{#2}}

\bibitem[{\citenamefont{Fraval et~al.}(2005)\citenamefont{Fraval, Sellars, and
  j.j. Longdell}}]{Fraval}
\bibinfo{author}{\bibfnamefont{E.}~\bibnamefont{Fraval}},
  \bibinfo{author}{\bibfnamefont{M.}~\bibnamefont{Sellars}}, \bibnamefont{and}
  \bibinfo{author}{\bibnamefont{j.j. Longdell}}, \bibinfo{journal}{Phys. Rev.
  Lett.} \textbf{\bibinfo{volume}{95}}, \bibinfo{pages}{030506}
  (\bibinfo{year}{2005}).

\bibitem[{\citenamefont{Jing and Mukamel}(1988)}]{Mukamel1}
\bibinfo{author}{\bibfnamefont{Y.}~\bibnamefont{Jing}} \bibnamefont{and}
  \bibinfo{author}{\bibfnamefont{S.}~\bibnamefont{Mukamel}},
  \bibinfo{journal}{J. Chem. Phys.} \textbf{\bibinfo{volume}{89}},
  \bibinfo{pages}{5160} (\bibinfo{year}{1988}).

\bibitem[{\citenamefont{Mukamel et~al.}(1988)\citenamefont{Mukamel, Tretiak,
  Wagerstreiter, and Chernyak}}]{Mukamel2}
\bibinfo{author}{\bibfnamefont{S.}~\bibnamefont{Mukamel}},
  \bibinfo{author}{\bibfnamefont{S.}~\bibnamefont{Tretiak}},
  \bibinfo{author}{\bibfnamefont{T.}~\bibnamefont{Wagerstreiter}},
  \bibnamefont{and} \bibinfo{author}{\bibfnamefont{V.}~\bibnamefont{Chernyak}},
  \bibinfo{journal}{J. Chem. Phys.} \textbf{\bibinfo{volume}{89}},
  \bibinfo{pages}{5160} (\bibinfo{year}{1988}).

\bibitem[{\citenamefont{Kohen et~al.}(1997)\citenamefont{Kohen, Marston, and
  Tannor}}]{tannorDC1}
\bibinfo{author}{\bibfnamefont{D.}~\bibnamefont{Kohen}},
  \bibinfo{author}{\bibfnamefont{C.~C.} \bibnamefont{Marston}},
  \bibnamefont{and} \bibinfo{author}{\bibfnamefont{D.}~\bibnamefont{Tannor}},
  \bibinfo{journal}{J. Chem. Phys.} \textbf{\bibinfo{volume}{107}},
  \bibinfo{pages}{5236} (\bibinfo{year}{1997}).

\bibitem[{\citenamefont{Kohen and Tannor}(1997)}]{tannorDC2}
\bibinfo{author}{\bibfnamefont{D.}~\bibnamefont{Kohen}} \bibnamefont{and}
  \bibinfo{author}{\bibfnamefont{D.~J.} \bibnamefont{Tannor}},
  \bibinfo{journal}{J. Chem. Phys.} \textbf{\bibinfo{volume}{107}},
  \bibinfo{pages}{5141} (\bibinfo{year}{1997}).

\bibitem[{\citenamefont{Bittner and Rossky}(1995)}]{Rossky1}
\bibinfo{author}{\bibfnamefont{E.}~\bibnamefont{Bittner}} \bibnamefont{and}
  \bibinfo{author}{\bibfnamefont{P.}~\bibnamefont{Rossky}},
  \bibinfo{journal}{J. Chem. Phys.} \textbf{\bibinfo{volume}{103}},
  \bibinfo{pages}{8130} (\bibinfo{year}{1995}).

\bibitem[{\citenamefont{Prezhdo and Rossky}(1998)}]{Rossky2}
\bibinfo{author}{\bibfnamefont{O.}~\bibnamefont{Prezhdo}} \bibnamefont{and}
  \bibinfo{author}{\bibfnamefont{P.}~\bibnamefont{Rossky}},
  \bibinfo{journal}{Phys. Rev. Lett.} \textbf{\bibinfo{volume}{81}},
  \bibinfo{pages}{5294} (\bibinfo{year}{1998}).

\bibitem[{\citenamefont{Karkuszewski et~al.}(2002)\citenamefont{Karkuszewski,
  Jarzynski, and Zurek}}]{Zurek1}
\bibinfo{author}{\bibfnamefont{Z.~P.} \bibnamefont{Karkuszewski}},
  \bibinfo{author}{\bibfnamefont{C.}~\bibnamefont{Jarzynski}},
  \bibnamefont{and} \bibinfo{author}{\bibfnamefont{W.}~\bibnamefont{Zurek}},
  \bibinfo{journal}{Phys. Rev. Lett.} \textbf{\bibinfo{volume}{89}},
  \bibinfo{pages}{170405} (\bibinfo{year}{2002}).

\bibitem[{\citenamefont{Zurek et~al.}(1993)\citenamefont{Zurek, Habib, and
  Paz}}]{Zurek2}
\bibinfo{author}{\bibfnamefont{W.}~\bibnamefont{Zurek}},
  \bibinfo{author}{\bibfnamefont{S.}~\bibnamefont{Habib}}, \bibnamefont{and}
  \bibinfo{author}{\bibfnamefont{J.}~\bibnamefont{Paz}},
  \bibinfo{journal}{Phys. Rev. Lett.} \textbf{\bibinfo{volume}{70}},
  \bibinfo{pages}{1187} (\bibinfo{year}{1993}).

\bibitem[{\citenamefont{Zurek}(1991)}]{Zurek3}
\bibinfo{author}{\bibfnamefont{W.}~\bibnamefont{Zurek}},
  \bibinfo{journal}{"PHYSICS TODAY"} \textbf{\bibinfo{volume}{44}},
  \bibinfo{pages}{36} (\bibinfo{year}{1991}).

\bibitem[{\citenamefont{{W. Zhu and M. Smit and H. Rabitz}}(1999)}]{rabitz4}
\bibinfo{author}{\bibnamefont{{W. Zhu and M. Smit and H. Rabitz}}},
  \bibinfo{journal}{J. Chem. Phys.} \textbf{\bibinfo{volume}{110}},
  \bibinfo{pages}{1905} (\bibinfo{year}{1999}).

\bibitem[{\citenamefont{{W. Zhu and H. Rabitz}}(2003)}]{rabitz6}
\bibinfo{author}{\bibnamefont{{W. Zhu and H. Rabitz}}}, \bibinfo{journal}{J.
  Chem. Phys.} \textbf{\bibinfo{volume}{119}}, \bibinfo{pages}{3619}
  (\bibinfo{year}{2003}).

\bibitem[{\citenamefont{Guhr and Schwentner}(2005)}]{schwentner1}
\bibinfo{author}{\bibfnamefont{M.}~\bibnamefont{Guhr}} \bibnamefont{and}
  \bibinfo{author}{\bibfnamefont{N.}~\bibnamefont{Schwentner}},
  \bibinfo{journal}{Phys. Chem. Chem. Phys} \textbf{\bibinfo{volume}{7}},
  \bibinfo{pages}{760} (\bibinfo{year}{2005}).

\bibitem[{\citenamefont{Guhr et~al.}(2004)\citenamefont{Guhr, Ibrahim, and
  Schwentner}}]{schwentner2}
\bibinfo{author}{\bibfnamefont{M.}~\bibnamefont{Guhr}},
  \bibinfo{author}{\bibfnamefont{H.}~\bibnamefont{Ibrahim}}, \bibnamefont{and}
  \bibinfo{author}{\bibfnamefont{N.}~\bibnamefont{Schwentner}},
  \bibinfo{journal}{Phys. Chem. Chem. Phys} \textbf{\bibinfo{volume}{6}},
  \bibinfo{pages}{5353} (\bibinfo{year}{2004}).

\bibitem[{\citenamefont{Lindblad}(1976)}]{lindblad}
\bibinfo{author}{\bibfnamefont{G.}~\bibnamefont{Lindblad}},
  \bibinfo{journal}{Commun. Math. Phys.} \textbf{\bibinfo{volume}{48}},
  \bibinfo{pages}{119} (\bibinfo{year}{1976}).

\bibitem[{\citenamefont{K\"{o}ppel et~al.}(1984)\citenamefont{K\"{o}ppel,
  Domcke, and Cederbaum}}]{koppel84}
\bibinfo{author}{\bibfnamefont{H.}~\bibnamefont{K\"{o}ppel}},
  \bibinfo{author}{\bibfnamefont{W.}~\bibnamefont{Domcke}}, \bibnamefont{and}
  \bibinfo{author}{\bibfnamefont{L.}~\bibnamefont{Cederbaum}},
  \bibinfo{journal}{Adv. Chem. Phys.} \textbf{\bibinfo{volume}{57}},
  \bibinfo{pages}{59} (\bibinfo{year}{1984}).

\bibitem[{\citenamefont{{W. Huisinga, L. Pesce, R. Kosloff and P.
  Saalfrank}}(1999)}]{k144}
\bibinfo{author}{\bibnamefont{{W. Huisinga, L. Pesce, R. Kosloff and P.
  Saalfrank}}}, \bibinfo{journal}{J. Chem. Phys.}
  \textbf{\bibinfo{volume}{110}}, \bibinfo{pages}{5538} (\bibinfo{year}{1999}).

\bibitem[{\citenamefont{{R. Kosloff and H. Tal-Ezer}}(1986)}]{k42}
\bibinfo{author}{\bibnamefont{{R. Kosloff and H. Tal-Ezer}}},
  \bibinfo{journal}{Chem. Phys. Lett.} \textbf{\bibinfo{volume}{127}},
  \bibinfo{pages}{223} (\bibinfo{year}{1986}).

\bibitem[{\citenamefont{{A. Bartana, R. Kosloff and D. J.
  Tannor}}(2001)}]{k163}
\bibinfo{author}{\bibnamefont{{A. Bartana, R. Kosloff and D. J. Tannor}}},
  \bibinfo{journal}{Chem. Phys.} \textbf{\bibinfo{volume}{267}},
  \bibinfo{pages}{195} (\bibinfo{year}{2001}).

\bibitem[{\citenamefont{Ketterle and Pritchard}(1992)}]{pritchard}
\bibinfo{author}{\bibfnamefont{W.}~\bibnamefont{Ketterle}} \bibnamefont{and}
  \bibinfo{author}{\bibfnamefont{D.~E.} \bibnamefont{Pritchard}},
  \bibinfo{journal}{Phys. Rev. A} \textbf{\bibinfo{volume}{46}},
  \bibinfo{pages}{4051} (\bibinfo{year}{1992}).

\bibitem[{\citenamefont{{D.A. Steck, K. Jacobs, H. Mabuchi, T. Bhattacharya and
  S. Habib}}(2004)}]{habib}
\bibinfo{author}{\bibnamefont{{D.A. Steck, K. Jacobs, H. Mabuchi, T.
  Bhattacharya and S. Habib}}}, \bibinfo{journal}{Phys. Rev. Lett.}
  \textbf{\bibinfo{volume}{92}}, \bibinfo{pages}{223004}
  (\bibinfo{year}{2004}).

\end{thebibliography}

\end{document}